\pdfoutput=1

\documentclass[12pt,a4paper]{article}

\usepackage{ifthen} 
\usepackage{subcaption}
\usepackage{multirow}

\usepackage{afterpage}

\newboolean{pdflatex}
\setboolean{pdflatex}{true} 

\newboolean{articletitles}
\setboolean{articletitles}{true} 

\newboolean{uprightparticles}
\setboolean{uprightparticles}{false} 

\newboolean{inbibliography}
\setboolean{inbibliography}{false} 

\usepackage{microtype}
\usepackage{lineno}  
\usepackage{xspace} 
\usepackage{caption} 

\usepackage{graphicx}  
\usepackage{color}
\usepackage{colortbl}
\graphicspath{{./figs/}} 

\usepackage{amsmath} 
\usepackage{amssymb}
\usepackage{amsfonts}
\usepackage{upgreek} 

\newcommand*\patchAmsMathEnvironmentForLineno[1]{%
\expandafter\let\csname old#1\expandafter\endcsname\csname #1\endcsname
\expandafter\let\csname oldend#1\expandafter\endcsname\csname
end#1\endcsname
 \renewenvironment{#1}%
   {\linenomath\csname old#1\endcsname}%
   {\csname oldend#1\endcsname\endlinenomath}%
}
\newcommand*\patchBothAmsMathEnvironmentsForLineno[1]{%
  \patchAmsMathEnvironmentForLineno{#1}%
  \patchAmsMathEnvironmentForLineno{#1*}%
}
\AtBeginDocument{%
\patchBothAmsMathEnvironmentsForLineno{equation}%
\patchBothAmsMathEnvironmentsForLineno{align}%
\patchBothAmsMathEnvironmentsForLineno{flalign}%
\patchBothAmsMathEnvironmentsForLineno{alignat}%
\patchBothAmsMathEnvironmentsForLineno{gather}%
\patchBothAmsMathEnvironmentsForLineno{multline}%
\patchBothAmsMathEnvironmentsForLineno{eqnarray}%
}

\usepackage{hyperref}    
\usepackage[all]{hypcap} 

\def\lhcb {\mbox{LHCb}\xspace}

\def\MagUp {\mbox{\em Mag\kern -0.05em Up}\xspace}

\def\lone   {L0\xspace}

\def\hltone {HLT1\xspace}
\def\hlttwo {HLT2\xspace}

\ifthenelse{\boolean{uprightparticles}}%
{

 \def\Pmu         {\ensuremath{\upmu}\xspace}

 \def\Ppi         {\ensuremath{\uppi}\xspace}

 \def\Ppsi        {\ensuremath{\uppsi}\xspace}

 \def\PDelta      {\ensuremath{\Delta}\xspace}                 
 \def\PXi      {\ensuremath{\Xi}\xspace}                 
 \def\PLambda      {\ensuremath{\Lambda}\xspace}                 
 \def\PSigma      {\ensuremath{\Sigma}\xspace}                 
 \def\POmega      {\ensuremath{\Omega}\xspace}                 
 \def\PUpsilon      {\ensuremath{\Upsilon}\xspace}                 
 
 \def\PB      {\ensuremath{\mathrm{B}}\xspace}                 
                  
 \def\PD      {\ensuremath{\mathrm{D}}\xspace}

 \def\PJ      {\ensuremath{\mathrm{J}}\xspace}                 
 \def\PK      {\ensuremath{\mathrm{K}}\xspace}

 \def\PZ      {\ensuremath{\mathrm{Z}}\xspace}                 
                  
 \def\Pb      {\ensuremath{\mathrm{b}}\xspace}                 
 \def\Pc      {\ensuremath{\mathrm{c}}\xspace}                 
 \def\Pd      {\ensuremath{\mathrm{d}}\xspace}

 \def\Pi      {\ensuremath{\mathrm{i}}\xspace}

 \def\Pq      {\ensuremath{\mathrm{q}}\xspace}                 
                  
 \def\Ps      {\ensuremath{\mathrm{s}}\xspace}                 
                  
 \def\Pu      {\ensuremath{\mathrm{u}}\xspace}

}
{

 \def\Pmu         {\ensuremath{\mu}\xspace}

 \def\Ppi         {\ensuremath{\pi}\xspace}

 \def\Ppsi        {\ensuremath{\psi}\xspace}                 
                  
 \mathchardef\PDelta="7101
 \mathchardef\PXi="7104
 \mathchardef\PLambda="7103
 \mathchardef\PSigma="7106
 \mathchardef\POmega="710A
 \mathchardef\PUpsilon="7107
                  
 \def\PB      {\ensuremath{B}\xspace}                 
                  
 \def\PD      {\ensuremath{D}\xspace}

 \def\PJ      {\ensuremath{J}\xspace}                 
 \def\PK      {\ensuremath{K}\xspace}

 \def\PZ      {\ensuremath{Z}\xspace}                 
                  
 \def\Pb      {\ensuremath{b}\xspace}                 
 \def\Pc      {\ensuremath{c}\xspace}                 
 \def\Pd      {\ensuremath{d}\xspace}

 \def\Pi      {\ensuremath{i}\xspace}

 \def\Pq      {\ensuremath{q}\xspace}                 
                  
 \def\Ps      {\ensuremath{s}\xspace}                 
                  
 \def\Pu      {\ensuremath{u}\xspace}

}

\makeatletter
\ifcase \@ptsize \relax
  \newcommand{\miniscule}{\@setfontsize\miniscule{4}{5}}
\or
  \newcommand{\miniscule}{\@setfontsize\miniscule{5}{6}}
\or
  \newcommand{\miniscule}{\@setfontsize\miniscule{5}{6}}
\fi
\makeatother

\DeclareRobustCommand{\optbar}[1]{\shortstack{{\miniscule (\rule[.5ex]{1.25em}{.18mm})}
  \\ [-.7ex] $#1$}}

\def\mup        {{\ensuremath{\Pmu^+}}\xspace}
\def\mun        {{\ensuremath{\Pmu^-}}\xspace} 

\def\quark     {{\ensuremath{\Pq}}\xspace}
\def\quarkbar  {{\ensuremath{\overline \quark}}\xspace}
\def\qqbar     {{\ensuremath{\quark\quarkbar}}\xspace}
\def\uquark    {{\ensuremath{\Pu}}\xspace}
\def\uquarkbar {{\ensuremath{\overline \uquark}}\xspace}
\def\uubar     {{\ensuremath{\uquark\uquarkbar}}\xspace}
\def\dquark    {{\ensuremath{\Pd}}\xspace}
\def\dquarkbar {{\ensuremath{\overline \dquark}}\xspace}
\def\ddbar     {{\ensuremath{\dquark\dquarkbar}}\xspace}
\def\squark    {{\ensuremath{\Ps}}\xspace}
\def\squarkbar {{\ensuremath{\overline \squark}}\xspace}
\def\ssbar     {{\ensuremath{\squark\squarkbar}}\xspace}
\def\cquark    {{\ensuremath{\Pc}}\xspace}
\def\cquarkbar {{\ensuremath{\overline \cquark}}\xspace}
\def\ccbar     {{\ensuremath{\cquark\cquarkbar}}\xspace}
\def\bquark    {{\ensuremath{\Pb}}\xspace}
\def\bquarkbar {{\ensuremath{\overline \bquark}}\xspace}
\def\bbbar     {{\ensuremath{\bquark\bquarkbar}}\xspace}

\def\pion   {{\ensuremath{\Ppi}}\xspace}

\def\pim    {{\ensuremath{\pion^-}}\xspace}

\def\kaon    {{\ensuremath{\PK}}\xspace}
  \def\Kbar    {{\kern 0.2em\overline{\kern -0.2em \PK}{}}\xspace}

\def\KorKbar    {\kern 0.18em\optbar{\kern -0.18em K}{}\xspace}

\def\Kp      {{\ensuremath{\kaon^+}}\xspace}

\def\KS      {{\ensuremath{\kaon^0_{\rm\scriptscriptstyle S}}}\xspace}

  \def\Dbar    {{\kern 0.2em\overline{\kern -0.2em \PD}{}}\xspace}

\def\DorDbar    {\kern 0.18em\optbar{\kern -0.18em D}{}\xspace}

\def\B       {{\ensuremath{\PB}}\xspace}
\def\Bbar    {{\ensuremath{\kern 0.18em\overline{\kern -0.18em \PB}{}}}\xspace}

\def\BorBbar    {\kern 0.18em\optbar{\kern -0.18em B}{}\xspace}

\def\Bd      {{\ensuremath{\B^0}}\xspace}

\def\jpsi     {{\ensuremath{{\PJ\mskip -3mu/\mskip -2mu\Ppsi\mskip 2mu}}}\xspace}

  \def\Y#1S{\ensuremath{\PUpsilon{(#1S)}}\xspace}

\def\Lbar        {{\ensuremath{\kern 0.1em\overline{\kern -0.1em\PLambda}}}\xspace}
\def\LorLbar    {\kern 0.18em\optbar{\kern -0.18em \PLambda}{}\xspace}

\def\to                 {\ensuremath{\rightarrow}\xspace}

\def\AT#1     {\ensuremath{A_{\mathrm{T}}^{#1}}\xspace}           

\def\C#1      {\ensuremath{\mathcal{C}_{#1}}\xspace}                       
\def\Cp#1     {\ensuremath{\mathcal{C}_{#1}^{'}}\xspace}                    
\def\Ceff#1   {\ensuremath{\mathcal{C}_{#1}^{\mathrm{(eff)}}}\xspace}        
\def\Cpeff#1  {\ensuremath{\mathcal{C}_{#1}^{'\mathrm{(eff)}}}\xspace}       
\def\Ope#1    {\ensuremath{\mathcal{O}_{#1}}\xspace}                       
\def\Opep#1   {\ensuremath{\mathcal{O}_{#1}^{'}}\xspace}                    

\newcommand{\tev}{\ifthenelse{\boolean{inbibliography}}{\ensuremath{~T\kern -0.05em eV}\xspace}{\ensuremath{\mathrm{\,Te\kern -0.1em V}}}\xspace}
\newcommand{\gev}{\ensuremath{\mathrm{\,Ge\kern -0.1em V}}\xspace}
\newcommand{\mev}{\ensuremath{\mathrm{\,Me\kern -0.1em V}}\xspace}
\newcommand{\kev}{\ensuremath{\mathrm{\,ke\kern -0.1em V}}\xspace}
\newcommand{\ev}{\ensuremath{\mathrm{\,e\kern -0.1em V}}\xspace}
\newcommand{\gevc}{\ensuremath{{\mathrm{\,Ge\kern -0.1em V\!/}c}}\xspace}
\newcommand{\mevc}{\ensuremath{{\mathrm{\,Me\kern -0.1em V\!/}c}}\xspace}
\newcommand{\gevcc}{\ensuremath{{\mathrm{\,Ge\kern -0.1em V\!/}c^2}}\xspace}
\newcommand{\gevgevcccc}{\ensuremath{{\mathrm{\,Ge\kern -0.1em V^2\!/}c^4}}\xspace}
\newcommand{\mevcc}{\ensuremath{{\mathrm{\,Me\kern -0.1em V\!/}c^2}}\xspace}

\def\mm   {\ensuremath{\rm \,mm}\xspace}

\def\mum  {\ensuremath{{\,\upmu\rm m}}\xspace}

\def\nb {\ensuremath{\rm \,nb}\xspace}

\def\pb {\ensuremath{\rm \,pb}\xspace}

\def\invfb   {\ensuremath{\mbox{\,fb}^{-1}}\xspace}

\def\ns   {\ensuremath{{\rm \,ns}}\xspace}
\def\ps   {\ensuremath{{\rm \,ps}}\xspace}

\def\gsim{{~\raise.15em\hbox{$>$}\kern-.85em
          \lower.35em\hbox{$\sim$}~}\xspace}
\def\lsim{{~\raise.15em\hbox{$<$}\kern-.85em
          \lower.35em\hbox{$\sim$}~}\xspace}

\def\ptot       {\mbox{$p$}\xspace}
\def\pt         {\mbox{$p_{\rm T}$}\xspace}

\def\evtgen     {\mbox{\textsc{EvtGen}}\xspace}

\def\geant      {\mbox{\textsc{Geant4}}\xspace}

\def\photos     {\mbox{\textsc{Photos}}\xspace}

\def\pythia     {\mbox{\textsc{Pythia}}\xspace}

\def\roofit     {\mbox{\textsc{RooFit}}\xspace}
\def\roostats     {\mbox{\textsc{RooStats}}\xspace}

\def\tell1  {TELL1\xspace}
\def\ukl1   {UKL1\xspace}

\usepackage{cite} 
\usepackage{mciteplus}

\newcommand{\vpion}{\ensuremath{{\pi_v}}\xspace}

\newcommand{\Rxy}{\ensuremath{R_{xy}}}
\newcommand{\mvtx}{\ensuremath{m_\text{vtx}}}
\newcommand{\Zplusjet}{\ensuremath{Z + \text{jet}}}
\newcommand{\percent}{\%}

\usepackage[separate-uncertainty=true]{siunitx}
\usepackage{graphicx}
\begin{document}

\renewcommand{\thefootnote}{\fnsymbol{footnote}}
\setcounter{footnote}{1}


\begin{titlepage}
\pagenumbering{roman}

\vspace*{-1.5cm}
\centerline{\large EUROPEAN ORGANIZATION FOR NUCLEAR RESEARCH (CERN)}
\vspace*{1.5cm}
\hspace*{-0.5cm}
\begin{tabular*}{\linewidth}{lc@{\extracolsep{\fill}}r}
\ifthenelse{\boolean{pdflatex}}
{\vspace*{-2.7cm}\mbox{\!\!\!\includegraphics[width=.14\textwidth]{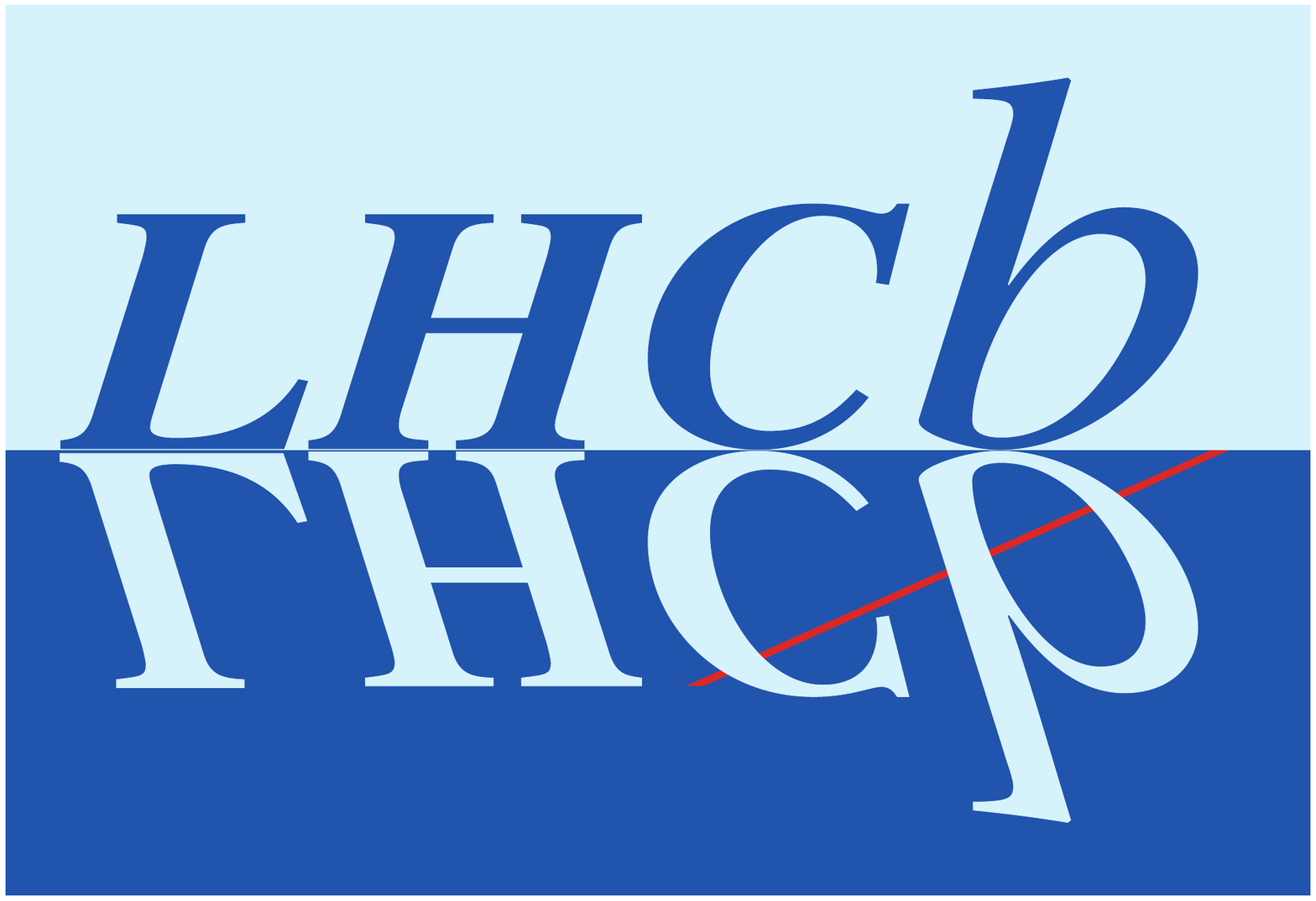}} & &}%
{\vspace*{-1.2cm}\mbox{\!\!\!\includegraphics[width=.12\textwidth]{lhcb-logo.eps}} & &}%
\\
 & & CERN-PH-EP-2014-291 \\  
 & & LHCb-PAPER-2014-062 \\  
 & & \today \\ 
 & & \\
\end{tabular*}

\vspace*{4.0cm}

{\bf\boldmath\huge
\begin{center}
  Search for long-lived particles decaying to jet pairs
\end{center}
}

\vspace*{2.0cm}

\begin{center}
The LHCb collaboration\footnote{Authors are listed at the end of this paper.}
\end{center}

\vspace{\fill}

\begin{abstract}
  \noindent
  A search is presented for long-lived particles with a mass between
  25 and 50\gevcc and a lifetime between 1 and
  200\ps in a sample of proton-proton collisions at a
  centre-of-mass energy of $\sqrt{s}=7\tev$, corresponding to
  an integrated luminosity of 0.62\invfb, collected by the
  \lhcb detector.  The particles are assumed to be pair-produced by
  the decay of a Standard Model-like Higgs boson. The experimental
  signature of the long-lived particle is a displaced vertex with two
  associated jets. No excess above the background is observed and
  limits are set on the production cross-section as a function of the
  long-lived particle mass and lifetime.
\end{abstract}

\vspace*{2.0cm}

\begin{center}
 Published as Eur.~Phys.~J.~C75 (2015) 152
\end{center}

\vspace{\fill}

{\footnotesize 
\centerline{\copyright~CERN on behalf of the \lhcb collaboration, license \href{http://creativecommons.org/licenses/by/4.0/}{CC-BY-4.0}.}}
\vspace*{2mm}

\end{titlepage}


\newpage
\setcounter{page}{2}
\mbox{~}

\cleardoublepage


\renewcommand{\thefootnote}{\arabic{footnote}}
\setcounter{footnote}{0}



\pagestyle{plain} 
\setcounter{page}{1}
\pagenumbering{arabic}


%

\section{Introduction}
\label{sec:Introduction}

A variety of models for physics beyond the Standard Model (SM) feature
the existence of new massive particles whose coupling to lighter
particles is sufficiently small that they are long-lived. If these
massive particles decay to SM particles and have a lifetime between
approximately 1\ps and 1\ns, characteristic of weak
decays, they can be identified by their
displaced decay vertex. Examples of such particles are the lightest
supersymmetric particle in SUSY models with baryon or lepton
number
violation~\cite{Carpenter:2007zz,*Butterworth:2009qa,*Kaplan:2007ap,*deCampos:2008re},
the next-to-lightest supersymmetric particle in gravity
mediated SUSY~\cite{deCampos:2013hca} and the neutral \vpion{}
particle in hidden valley (HV) models with a non-abelian gauge
symmetry~\cite{Strassler:2006im,*Strassler:2006ri,*Han:2007ae}.  The
latter model is particularly interesting as it predicts that
experimental studies have sensitivity to the production of long-lived particles in SM Higgs decays.

This paper reports on a search for \vpion{} particles, pair-produced
in the decay of a SM-like Higgs particle with a mass of
120\gevcc, close to the mass of the scalar boson discovered by
the ATLAS and CMS
experiments~\cite{Aad:2012tfa,Chatrchyan:2012ufa}.\footnote{The
  results are equally valid for a Higgs particle with a mass up to
  126\gevcc within a few percent.} The \vpion{}
candidates are identified by two hadronic jets originating from a
displaced vertex.
The vertex is required to be displaced from the proton-proton
collision axis by more than 0.4~mm and less than 4.8~mm. The lower
bound is chosen to reject most of the background from heavy flavour
decays. 
The upper bound ensures that vertices are inside the LHCb beam pipe,
which generates a sizeable background of hadronic interaction
vertices.
The signal is extracted from a fit to the di-jet
invariant mass distribution. The analysis is sensitive to a \vpion{}
particle with a mass between 25 and 50\gevcc and a
lifetime between 1 and 200\ps. The lower boundary on the mass
range arises from the requirement to identify two hadronic jets while
the upper boundary is mostly due to the geometric acceptance of the
LHCb detector.

This analysis uses data collected in proton-proton ($pp$) collisions
at a centre-of-mass energy of $\sqrt{s}=7\tev$. The data
correspond to an integrated luminosity of 0.62\invfb, collected
during the second half of the year 2011 when an analysis-specific
trigger selection was implemented. Although similar searches have been reported
by the CDF~\cite{Aaltonen:2011rja}, D0~\cite{Abazov:2009ik},
ATLAS~\cite{ATLAS:2012av} and CMS~\cite{CMS:2014wda} experiments, LHCb has a unique coverage for
long-lived particles with relatively small mass and lifetime, because
its trigger makes only modest requirements on transverse momentum.

\section{Detector description}
\label{sec:Detector}
The \lhcb detector~\cite{Alves:2008zz} is a single-arm forward
spectrometer covering the \mbox{pseudorapidity} range $2<\eta <5$,
designed for the study of particles containing \bquark or \cquark
quarks. The detector includes a high-precision tracking system
consisting of a silicon-strip vertex detector surrounding the $pp$
interaction region~\cite{LHCb-DP-2014-001}, a large-area silicon-strip
detector located upstream of a dipole magnet with a bending power of
about 4\,Tm, and three stations of silicon-strip detectors and straw
drift tubes~\cite{LHCb-DP-2013-003} placed downstream of the magnet.
The tracking system provides a measurement of momentum, \ptot, with a
relative uncertainty that varies from \SI{0.4}{\percent} at low
momentum to \SI{0.6}{\percent} at 100\gevc.  The minimum
distance of a track to a primary vertex, the impact parameter, is
measured with a resolution of $(15+29/\pt)\mum$, where \pt is the
component of \ptot transverse to the beam, in \gevc.
Different types of charged hadrons are distinguished using information
from two ring-imaging Cherenkov
detectors~\cite{LHCb-DP-2012-003}. Photon, electron and hadron
candidates are identified by a calorimeter system consisting of
scintillating-pad and preshower detectors, an electromagnetic
calorimeter and a hadronic calorimeter. Muons are identified by a
system composed of alternating layers of iron and multiwire
proportional chambers~\cite{LHCb-DP-2012-002}.

\section{Event simulation}
\label{sec:Simulation}

For the event simulation, $pp$ collisions are generated using
\pythia6.4~\cite{Sjostrand:2006za} with a specific \lhcb
configuration~\cite{LHCb-PROC-2010-056} using CTEQ6L~\cite{cteq6l} parton density functions.  Decays of hadronic particles
are described by \evtgen~\cite{Lange:2001uf}, in which final-state
radiation is generated using \photos~\cite{Golonka:2005pn}. The
interaction of the generated particles with the detector and its
response are implemented using the \geant
toolkit~\cite{Allison:2006ve, *Agostinelli:2002hh} as described in
Ref.~\cite{LHCb-PROC-2011-006}.

To simulate a signal event, a SM-like scalar Higgs boson with a mass of 120\gevcc is
generated with \pythia through the gluon-gluon fusion mechanism, and is forced to decay into two spin-zero \vpion
particles, each of which decays to \bbbar. Assuming the decay occurs
via a vector or axial-vector coupling, the \bbbar final state is
preferred to light quarks, due to helicity
conservation~\cite{Strassler:2006im,*Strassler:2006ri,*Han:2007ae}.
The average track multiplicity of the \vpion decay, including tracks
from secondary $b$ and $c$ decays, varies from about 15 for a
\vpion mass of 25\gevcc to about 20 for larger masses.
Simulated events are retained if at least four charged tracks from the decay
of the generated \vpion particles are within the \lhcb acceptance, which
corresponds to about \SI{30}{\percent} of the cases. For \vpion
particles within the acceptance
on average about ten tracks can be reconstructed.

Simulated samples with \vpion lifetimes of 10\ps and
100\ps and \vpion masses of 25, 35, 43 and
50\gevcc are generated; other \vpion lifetimes are studied by reweighting these
samples. Two additional samples
are generated in which \vpion particles with a lifetime of
10\ps and a mass of 35\gevcc decay
to either \ccbar or \ssbar quark pairs.

\section{Event selection and signal extraction}
\label{sec:selection}

The selection of candidates starts with the LHCb
trigger~\cite{LHCb-DP-2012-004}, which consists of a hardware stage,
based on information from the calorimeter and muon systems, followed
by a software stage, which applies a full event reconstruction. The
hardware trigger (\lone) requires a single high-\pt{}
hadron, electron, muon or photon signature. The thresholds range from
$\pt{}>1.48\gevc$ for muons, to transverse energy larger than $3.5\gev$ for hadrons. The total \lone efficiency, dominated
by the hadron trigger selection, depends on the mass and final state
of the \vpion{} particle and is typically \SI{20}{\percent}, including
the detector acceptance.

The software trigger is divided into two stages and consists of
algorithms that run a simplified version of the offline track
reconstruction, which allows identification of displaced
tracks and vertices. For this analysis the primary signature in the
first software stage (\hltone) is a single high-quality displaced
track with high \pt{}. The efficiency of \hltone relative
to \lone accepted events is typically
\SI{60}{\percent}. However, this efficiency reduces rapidly for
vertices that are displaced by more than about 5\mm from the
beamline due to limitations
in the track reconstruction in the vertex detector.

In the final trigger stage (\hlttwo) two different signatures are
exploited. The first of these relies on the generic reconstruction of a
displaced vertex, using an algorithm similar to that used for
the primary vertex (PV) reconstruction~\cite{Kucharczyk:1756296}. Secondary vertices are distinguished
from PVs using the distance to the interaction region in the
transverse plane (\Rxy{}). To eliminate contributions from
interactions with material, a so-called `material veto' removes
vertices in a region defined as an envelope around the
detector material~\cite{LHCb-PAPER-2012-023}. Events are selected when
they have a displaced vertex with at least four tracks, a
sum of the scalar \pt{} of all tracks that is larger than 3\gevc, a distance \Rxy{} larger
than 0.4\mm and an invariant mass of the particles associated
with this vertex \mvtx{} above
4.5\gevcc. To further refine the selection, vertices
are required to have either $\Rxy > 2\mm$ or $\mvtx > 10\gevcc$.

The second \hlttwo signature is designed to identify two-,
three- and four-body exclusive $b$-hadron
decays~\cite{BBDT}. 
A multivariate
algorithm is used for the identification of secondary
vertices consistent with the decay of a \bquark{} hadron.  The
combined efficiency of the two \hlttwo selections relative to events
accepted by \lone and
\hltone is about \SI{60}{\percent}.

The offline candidate reconstruction starts from a generic secondary
vertex search, similar to that applied in the trigger, but using 
tracks from the offline reconstruction as input. At this stage at least six tracks per
vertex are required and the sum of the scalar \pt{} of all tracks must be above
3\gevc.
The vertex is required to have either $\Rxy>0.4\mm$ and $\mvtx >
9.7\gevcc$, or $\Rxy>2.5\mm$ and $\mvtx > 8.5\gevcc$,
or $\Rxy > 4\mm$ and $\mvtx > 6.5\gevcc$.

The vertex reconstruction is followed by a jet reconstruction
procedure. Inputs to the jet clustering are obtained using a
particle flow approach~\cite{Aaij:2013nxa} that selects charged
particles, neutral calorimeter deposits and a small contribution from \KS{} and $\PLambda^0$ decays. To
reduce contamination from particles that do not originate from the
displaced vertex, only charged particles that have a smaller distance
of closest approach
relative to the displaced vertex than to any PV in the event are
retained. Furthermore, the distance to the
displaced vertex is required to be less than 2\mm, which also
allows tracks from displaced $b$ and $c$
vertices in the $\vpion\to\bbbar$ decay chain to be accepted.

The jet clustering uses the anti-$k_\mathrm{T}$ algorithm~\cite{Cacciari:2008gp} with a cone size of
0.7. Only jets with a \pt{} above 5\gevc are used.
Additional requirements are made to enhance the fraction of
well-reconstructed hadronic jets: first, the charged particle with the
largest \pt{} in the jet must have a \pt{} above
0.9\gevc, yet carry no more than $\SI{70}{\percent}$ of the
\pt{} of the jet. 
Second, to remove jets whose energy is dominated by
neutral particles, which cannot be unambiguously associated with a
vertex, at least \SI{10}{\percent} of the \pt{} of the jet must be carried by charged particles.

The di-jet invariant mass is computed from the reconstructed
four-momenta of the two jets. Correction factors to the jet energy are
determined from the simulation and parameterised as a function of the
number of reconstructed PVs in the event, to account for effects due
to multiple interactions and the underlying event~\cite{Aaij:2013nxa}.

Two further requirements are made to
enhance signal purity. First, a corrected mass is computed as
\begin{linenomath} \begin{equation} m_\text{corr} \; = \; \sqrt{m^2 +
      \left(p \sin\theta\right)^2} \: + \: p \sin\theta \; ,
\end{equation}
\end{linenomath}
where $m$ is the di-jet invariant mass and $\theta$ is the pointing
angle between the di-jet momentum vector $\vec{p}$ and its displacement
vector $\vec{d} = \vec{x}_\text{DV} - \vec{x}_\text{PV}$, where
$\vec{x}_\text{DV}$ is the position of the displaced vertex and
$\vec{x}_\text{PV}$ the position of the PV. 
To select candidates pointing
back to a PV, only events with $m / m_\text{corr} > 0.7$ are
retained. A requirement on this ratio is preferred over a requirement
on the pointing angle itself, since its efficiency depends less
strongly on the boost and the mass of the candidate.

Second, a requirement is made on the distance $\Delta R =
\sqrt{\Delta\phi^2 + \Delta\eta^2} $ between the two jets, where $\phi$
is the azimuthal angle and $\eta$ the pseudorapidity. A
background consisting of back-to-back jet candidates, for example
di-jet \bbbar{}-events, appears mainly at large values of reconstructed
mass, and is characterised by
a large difference between the jets in both $\phi$ and $\eta$. Only
candidates with $\Delta R <
2.2$ are accepted.

\begin{table}[t]
  \caption{Average number of selected candidates per event
    (efficiency) in \percent{} for the main stages of the offline selection for simulated
    $H^0 \to \vpion \vpion$ events with $\vpion\to\bbbar$, $m_{H^0}
    =120\gevcc$, $m_{\vpion} = 35\gevcc$ and
    $\tau_{\vpion} =10\ps$. The pre-selection consists of the
    acceptance, trigger and offline vertex reconstruction. It
    represents the first stage in which the candidate yield on the total
    data sample, shown in the right column, can be counted. The
    reported uncertainty on the efficiency is only the statistical
    uncertainty from the finite sample size.}
  \label{tab:vpionefficiencyinsteps}
  \centerline{
    \begin{tabular}{lcr}
      \hline
      selection step                                  
      & signal efficiency & yield in data\\
      \hline
      pre-selection                     & \num{2.125 \pm 0.018}&2,555,377\\
      jet reconstruction                            & \num{1.207 \pm 0.014}&117,054\\
      $m/m_\text{corr}$  and $\Delta R$  & \num{0.873 \pm 0.012}&58,163\\
      trigger on candidate                        & \num{0.778 \pm 0.012}& 29,921\\
      \hline
    \end{tabular}
  }
\end{table}

\begin{table}[t]
  \caption{Average number of selected candidates per event
    (efficiency) in \percent{} for different
    \vpion masses, lifetimes and decay modes. The reported uncertainty is only the statistical
    uncertainty from the finite sample size. No simulated samples were
    generated for the
    100\ps decay to light quarks.}
  \label{tab:vpionefficiency}
  \centerline{
    \begin{tabular}{cccc}
      \hline 
      & &  \multicolumn{2}{c} {signal efficiency} \\
      decay & $m_{\vpion}$ [\gevcc]  & $\tau_{\vpion} = 10\ps$ & $\tau_{\vpion} = 100\ps$  \\
      \hline
\rule{0pt}{3ex} $\vpion \to \bbbar$ & \num{25} & \num{0.373 \pm 0.008} & \num{0.0805 \pm 0.0019} \\
                                 & \num{35} & \num{0.778 \pm 0.012} & \num{0.181  \pm 0.005 } \\
                                 & \num{43} & \num{0.743 \pm 0.011} & \num{0.183  \pm 0.003} \\
                                 & \num{50} & \num{0.573 \pm
                                   0.015} & \num{0.154  \pm 0.004}\\
\hline
$\vpion \to \ccbar$ & \num{35} & \num{2.18  \pm 0.05 } & --\\
$\vpion \to \ssbar$ & \num{35} & \num{2.06  \pm 0.04 } &--\\
      \hline
    \end{tabular}
  }
\end{table}

Finally, in order to facilitate a reliable estimate of the trigger
efficiency, only candidates triggered by particles belonging to one of
the jets are kept. Table~\ref{tab:vpionefficiencyinsteps} shows
the efficiency to select a \vpion particle, for an illustrative mass of 35\gevcc and
lifetime of 10\ps, together with the yield in the data after the most
important selection steps. The total efficiency for other masses and
lifetimes, as well as for the decays to light quark jets, is shown in
Table~\ref{tab:vpionefficiency}. The efficiencies listed in
Tables~\ref{tab:vpionefficiencyinsteps} and~\ref{tab:vpionefficiency} represent the number of selected
candidates divided by the number of generated events. As the
selection efficiencies for the two \vpion{} particles in an event are
practically independent, the fraction of selected events with more
than one candidate is less than a few percent in simulated signal. In data no events with more than one \vpion
candidate are found.

\begin{figure}[tb]
  \centerline{
\begin{subfigure}{0.49\linewidth}
  \includegraphics[width=\textwidth]{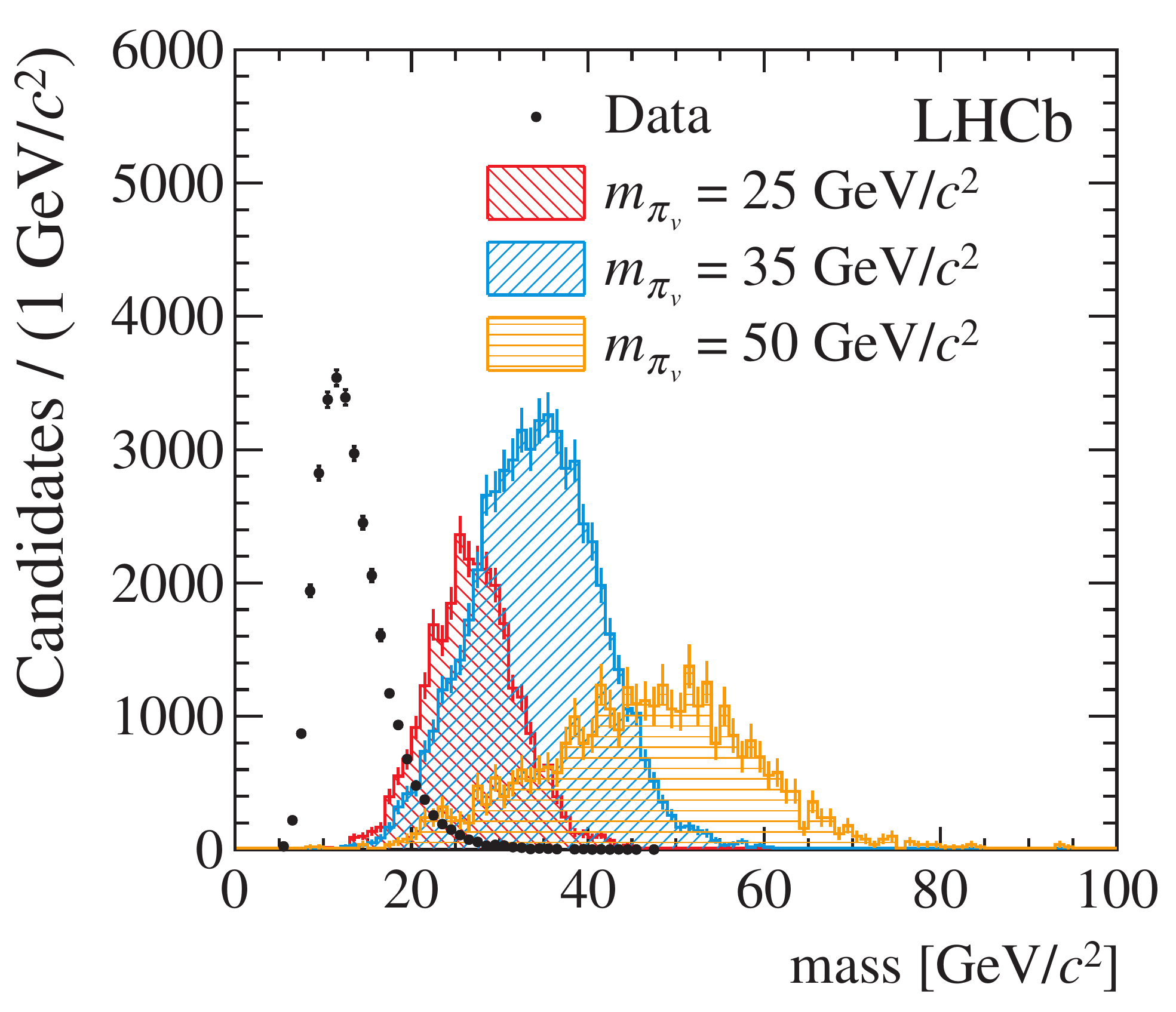}\llap{
  \parbox[b]{2.3in}{(a)\\\rule{0ex}{2.2in}}}
\phantomcaption\label{fig:massdistribution}
 \end{subfigure}
  \begin{subfigure}{0.49\linewidth}
\includegraphics[width=\textwidth]{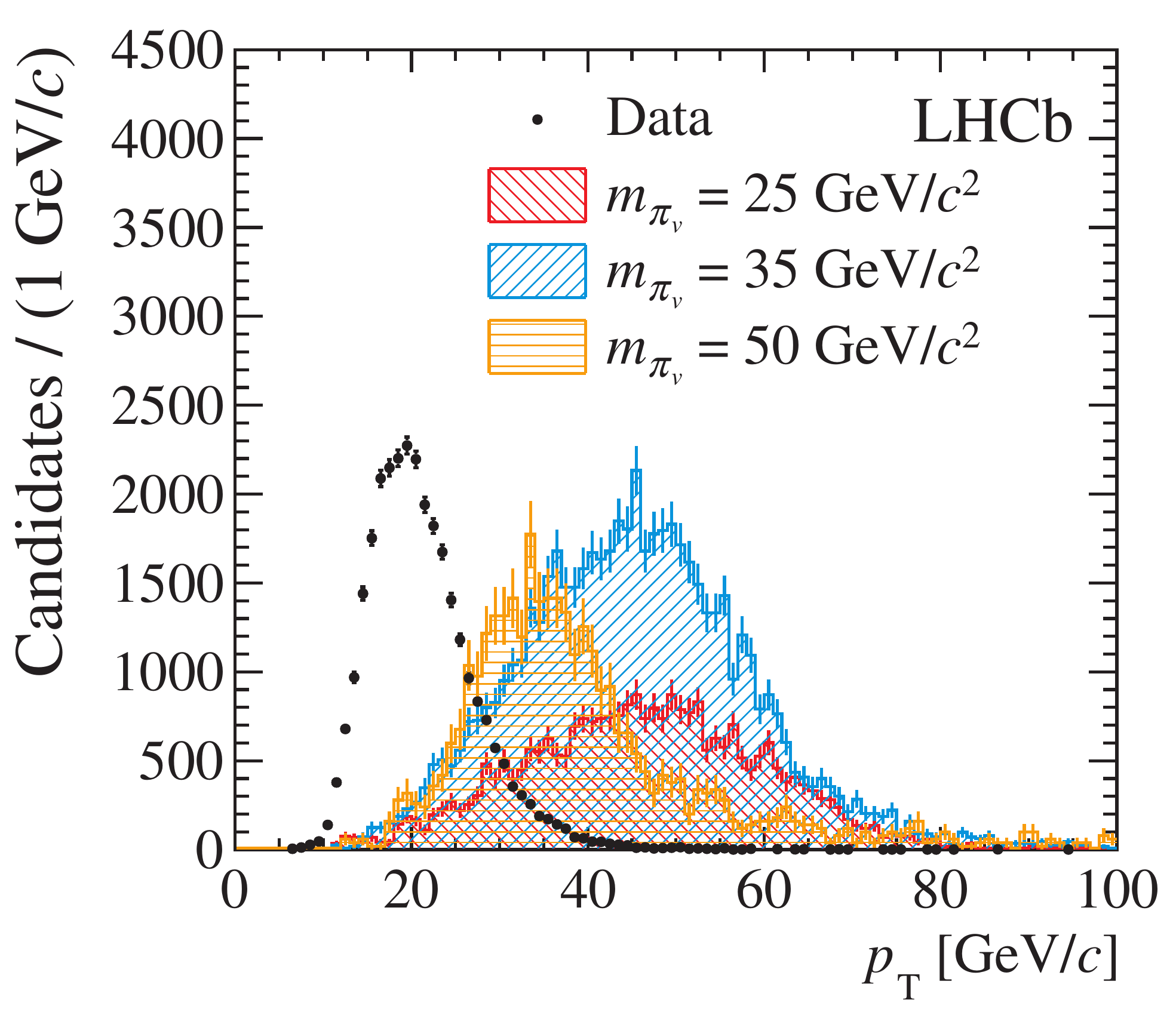}\llap{
  \parbox[b]{2.3in}{(b)\\\rule{0ex}{2.2in}}}
\phantomcaption\label{fig:ptdistribution}
 \end{subfigure}}
  \caption{Invariant mass (a) and \pt{} distribution (b) for di-jet
    candidates in data and in hidden
    valley models with \num{25}, \num{35} and 50\gevcc \vpion masses and 10\ps
    lifetime. For visibility, the simulated signal is scaled to
    0.62\invfb assuming a Higgs
    cross-section of 10\nb and branching fractions of
    \SI{100}{\percent} for
    ${\cal B}(H\to \vpion \vpion)$ and ${\cal B}(\vpion\to
    \bbbar)$.}
  \label{fig:massptdistribution}
\end{figure}

Figure~\ref{fig:massptdistribution} shows the mass and \pt
distributions for selected di-jet candidates in data and in 
simulated signal events, assuming a
\vpion{} particle with a mass of \num{25}, \num{35} or
50\gevcc. The turn-on at low values in the mass
distribution of events observed in data (Fig.~\ref{fig:massdistribution}) is caused
by the minimum \pt{} requirement on the jets. The rest of
the distribution falls off exponentially. 
The \pt distribution shown in
Fig.~\ref{fig:ptdistribution} illustrates that long-lived particles with a higher mass have
lower \pt as there is less momentum available in the Higgs decay. 
This affects the selection
efficiency since for a given decay time the transverse decay length is
proportional to \pt{}.

Studies on
simulated events have shown that both the shape and the normalisation of
the mass distribution in data are compatible with the expected
background from \bbbar{} production. It is not possible to generate
sufficiently large samples of \bbbar events to use these for a quantitative estimate of
the background after the final selection. Therefore, the signal yield is extracted by a fit to the invariant mass
distribution assuming a smooth shape for the background, as discussed
in Section~\ref{sec:Results}.

\begin{figure}[htb]
  \centering
  \includegraphics[width=0.5\textwidth]{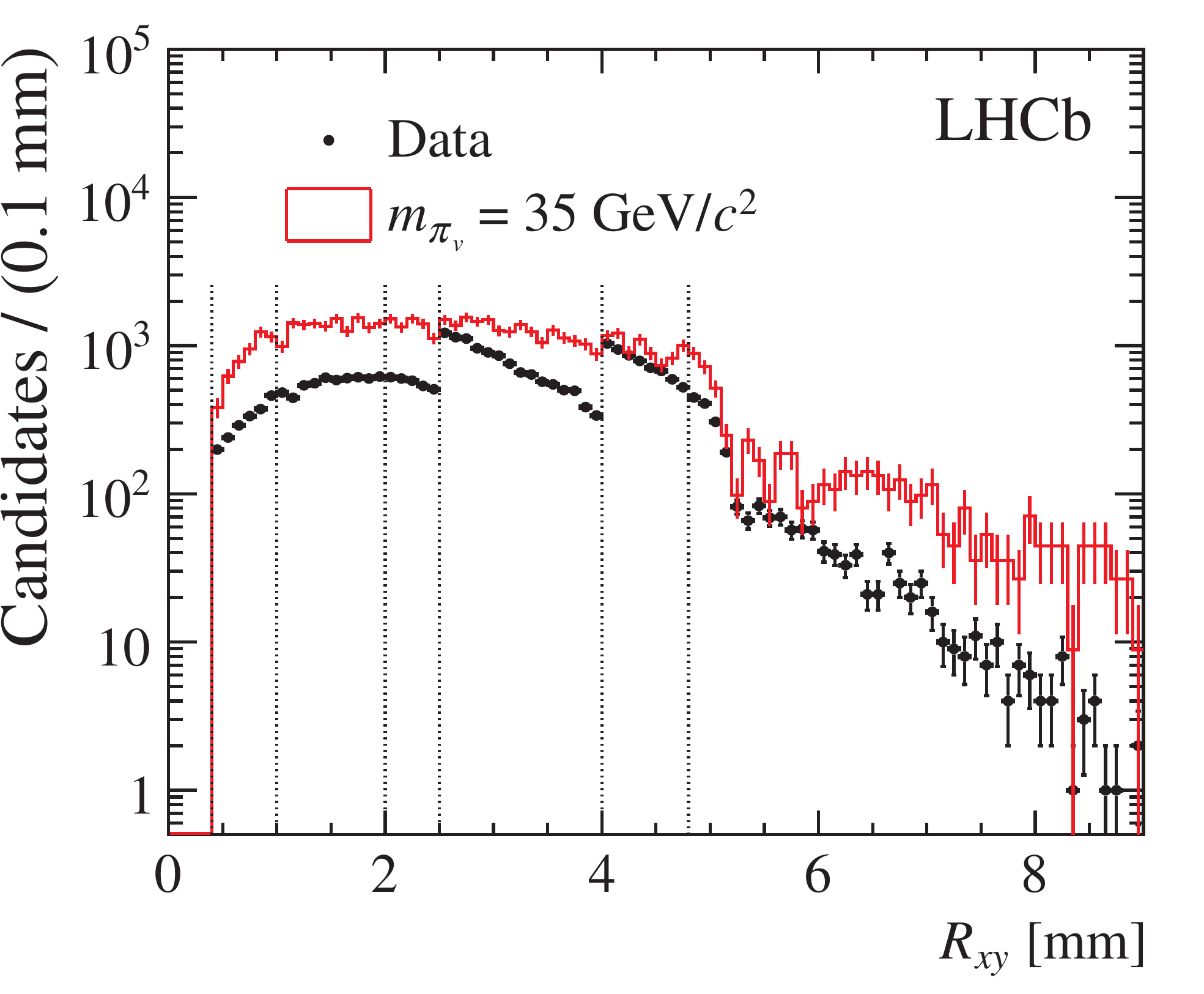}
  \caption{Distribution of the distance of the displaced
    vertex to the interaction region in the
    transverse plane for data and for a hidden valley model with $m_{\vpion} =
    35\gevcc$ and $\tau_{\vpion} =10\ps$ after the full
    selection. For visibility, the simulated signal is scaled to
    0.62\invfb assuming a Higgs
    cross-section of 10\nb and branching fractions of
    \SI{100}{\percent} for
    ${\cal B}(H\to \vpion \vpion)$ and ${\cal B}(\vpion\to
    \bbbar)$. The boundaries of the intervals used in the fit are
    indicated by the dotted lines. The generated \Rxy{}
    distribution is approximately exponential with an average of about
    2\mm.}
  \label{fig:rxydistribution}
\end{figure}

Since the background yield, the shape of the background invariant mass
distribution and the
selection efficiency strongly depend on the radial displacement
\Rxy{}, limits are extracted from a simultaneous maximum likelihood
fit to the di-jet invariant mass distribution in five bins of
\Rxy{}. The intervals are chosen in the most sensitive
region, between 0.4 and 4.8\mm. The events at larger radii
are not used as they contribute only marginally to the sensitivity. 
Figure~\ref{fig:rxydistribution} shows the distribution of \Rxy{} of selected
displaced vertices for data and simulated signal events, together with
the bin boundaries. The effect of the reduction in
efficiency at large radii due to the material veto and the HLT1
trigger is visible, as is the effect of requirements on \Rxy{} in the
trigger. The trigger effects are more pronounced in data than in
simulated signal, because signal events are less affected by cuts
on the vertex invariant mass.

\begin{figure}[h]
  \centering 
  \includegraphics[width=.9\linewidth]{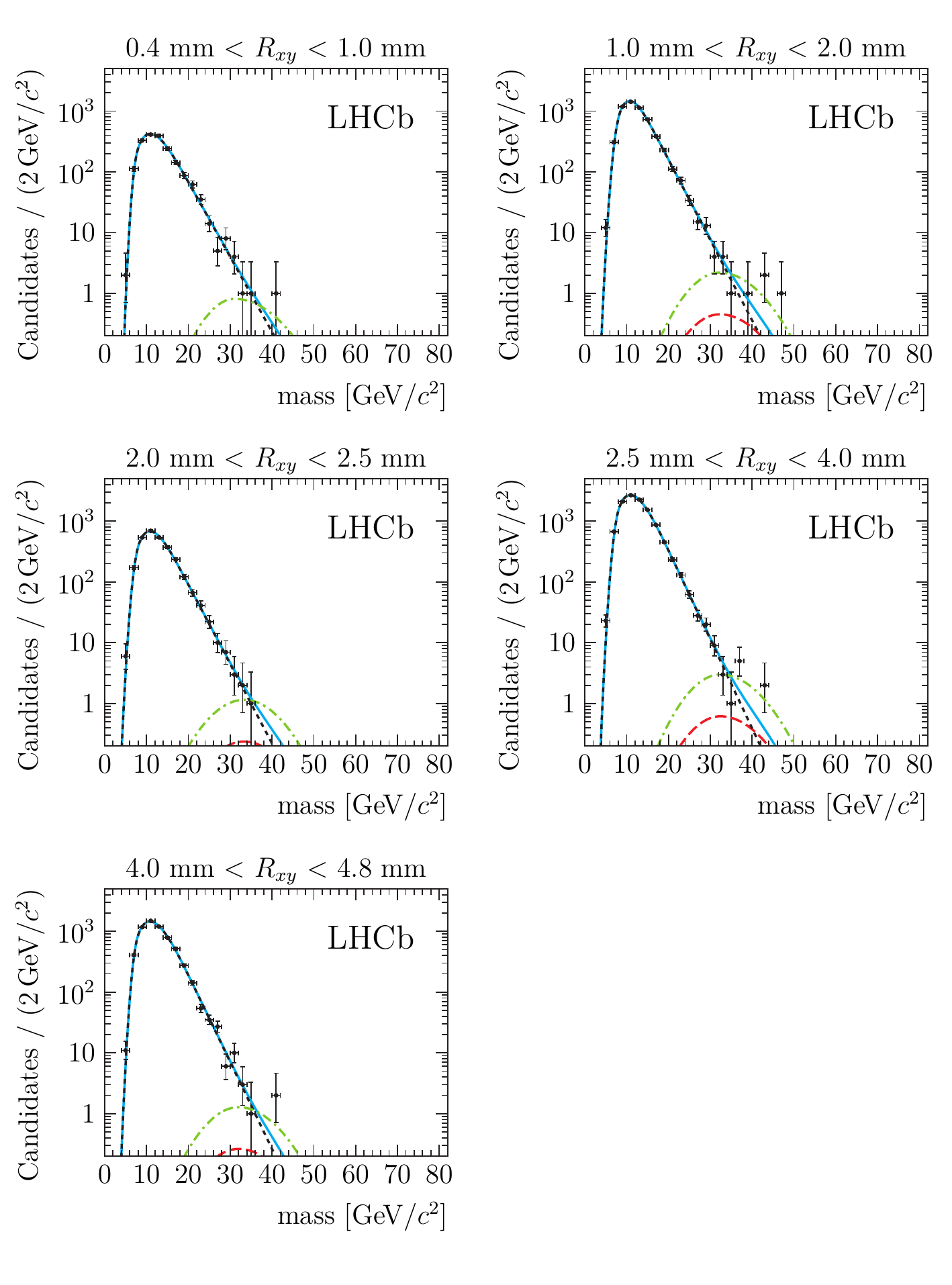}
  \caption{Di-jet invariant mass distributions for each of the five $\Rxy$ bins, superimposed with the
    fits for a hidden valley model with 
    $m_{\vpion} = 35\gevcc$ and $\tau_{\vpion}
    =10\ps$. The blue line indicates 
    the result of the total fit to the data. The black short-dashed line is the background-only contribution, and 
    the red long-dashed line is the fitted signal
    contribution.
    For illustration, the green dash-dotted line shows the signal scaled to a
    cross-section of 17\pb, which corresponds to the SM Higgs
    production cross-section at 7 TeV~\protect\cite{Heinemeyer:2013tqa}.}
  \label{fig:limit_bestfit35}
\end{figure}

The background di-jet invariant mass distribution is
characterised by an exponential falloff, with a low-mass threshold
determined mostly by the minimum \pt{} requirement of the jets.  It
is modelled by a single-sided exponential function convoluted with a
bifurcated Gaussian function. The parameters of the background model
are fitted to data, independently in each \Rxy{} bin. The signal is
modelled by a bifurcated Gaussian function, whose parameters are
determined from simulated events in bins of \Rxy{}. The effect of the
uncertainty on the jet-energy scale is included by a scale parameter
for the mass, which is common to all bins and constrained using a
sample of \Zplusjet{} events, as explained in
Section~\ref{sec:systematics}. Additional nuisance parameters are
added to account for the finite statistics of the simulated samples
and the systematic uncertainties on the signal efficiency and the
luminosity. The fit model is implemented using the
\roofit~\cite{Verkerke:2003ir}
package. Figure~\ref{fig:limit_bestfit35} shows the fit result in the
five radial bins for a signal model with $m_{\vpion} = 35\gevcc$ and
$\tau_{\vpion} = 10\ps$.

\afterpage{\clearpage}

\section{Systematic uncertainties}
\label{sec:systematics}

Several sources of systematic uncertainties have been considered. The
uncertainties depend on the \vpion{} mass and are summarised in
Table~\ref{tab:systematics}. The uncertainty on the vertex finding efficiency is assessed
by comparing the efficiency of the vertexing algorithm on a sample of
$\Bd\to\jpsi K^{*0}$ with $K^{*0}\to\Kp\pim$ events in data and
simulation as a function of
\Rxy{}.
The efficiency difference
is about \SI{7.5}{\percent} at large \Rxy{}, which is taken as an
estimate of the uncertainty on the vertex finding algorithm efficiency. Since the \Bd{} vertices
have only four tracks, and the \vpion decays
studied in this paper have typically more tracks, this is considered a
conservative estimate. The
uncertainty on the track finding efficiency for prompt tracks in LHCb
is \SI{1.4}{\percent} per track, with a small dependence on track
kinematics~\cite{LHCb-DP-2013-002}. The uncertainty for displaced tracks was evaluated in the
context of a recent \lhcb measurement of \mbox{\Pb-hadron} lifetimes~\cite{Aaij:2014owa} and
extrapolated to larger \Rxy, leading to a per-track uncertainty of
\SI{2}{\percent}. Due to requirements on the minimal number of tracks
in the vertex, this translates into an uncertainty on the vertex
finding efficiency, which is estimated to be \SI{2}{\percent} for
signal events. Adding in quadrature the track efficiency and
the vertex finding algorithm efficiency uncertainties leads to a total
uncertainty of \SI{7.9}{\percent} on the vertex reconstruction.
The selection on the vertex sum-\pt{} and mass is affected by the
track finding efficiency as well. Propagating the per-track
uncertainty leads to an uncertainty on the vertex selection efficiency
of up to \SI{2.9}{\percent}, depending on the \vpion{} mass.

The uncertainties related to the jet selection are determined by
comparing jets in data and simulation on a sample of \Zplusjet{}
events, analogously to a recent \lhcb measurement of \Zplusjet{}
production~\cite{Aaij:2013nxa}. The \PZ candidate is reconstructed in
the \mup\mun final state from two oppositely charged tracks,
identified as muons, that form a good vertex and have an invariant
mass in the range 60 -- 120\gev.  Jets are reconstructed
using the same selection of input particles as in the reconstruction
of jets for long-lived particles, except that the origin vertex is in
this case the PV consistent with the $Z$ vertex. The differences
between data and simulation in the \Zplusjet{} sample are
parameterised as function of the jet \pt{} and subsequently propagated
to the simulated hidden valley signal samples.

The uncertainty on the jet energy scale is derived from the ratio of 
transverse momenta of the jet and the $Z$, which are expected to have a back-to-back topology, and correlated transverse momenta. Data and simulation agree within about 
\SI{2}{\percent}, resulting in an uncertainty on the 
di-jet invariant mass scale of \SI{4}{\percent}. This uncertainty on 
the signal shape is taken into 
account in the fitting procedure.
The uncertainty on the jet-energy scale also affects the jet
reconstruction efficiency due to the requirement on the minimum jet
\pt{}. It leads to an uncertainty on the efficiency between 0.3 and \SI{1.3}{\percent}, depending on the assumed \vpion{} particle
mass.  The uncertainty on the hadronic jet identification requirements
are assessed using the \Zplusjet{} sample as well and amount to about
\SI{3}{\percent}.

The resolutions on the pointing angle $\theta$ and on $\Delta R$ are dominated by the
resolution on the direction of the \vpion{} candidate, which
in turn is determined by the jet angular resolution. The latter is estimated from the difference between data
and simulation in the
resolution of the azimuthal angle between the jet and
 the $Z$. Due to the limited statistics in the \Zplusjet{}
sample a relatively large uncertainty between 2.0 and
\SI{4.6}{\percent} is obtained, depending on the \vpion{} mass.

The trigger selection efficiency on signal is determined from
the simulation. The trigger efficiencies in data and simulation are
compared using a sample of generic $B\to J/\psi X$ events that contain
an offline reconstructed displaced vertex, but are triggered
independently of the displaced vertex trigger lines. The integrated
efficiency difference for the trigger stages \lone, \hltone and
\hlttwo amounts to systematic uncertainties of at most 4.6,
4.3 and \SI{6.3}{\percent} respectively. This is a conservative
estimate since the trigger efficiencies for the sample of displaced
$J/\psi$ vertices are smaller than the efficiencies for the signal,
which consists of heavier, more displaced objects with a larger number
of tracks. Finally, the uncertainty on the luminosity at the \lhcb interaction point is
\SI{1.7}{\percent}~\cite{LHCb-PAPER-2014-047}.

\begin{table}[t]
  \caption{Systematic uncertainties on the selection efficiency and
    luminosity for simulated hidden valley events with a lifetime of 10\ps and various \vpion masses.} 
  \label{tab:systematics}
  \centerline{
  \begin{tabular}{lrrrrr}
        \hline
        source & \multicolumn{5}{c} {relative uncertainty (\%)} \\
        \hline
      \rule{0pt}{3ex}  \vpion mass [\gevcc] & 25  & 35 & 43 & 50 \\
        \hline
        vertex reconstruction  &  7.9  &  7.9  &  7.9  &  7.9   \\
        vertex scalar-\pt{} and mass&  2.9  &  2.3  &  2.0  &  1.7   \\
        jet reconstruction  &  1.3  &  0.6  &  0.4  &  0.3   \\
        jet identification  &  2.9  &  3.0  &  3.2  &  3.2  \\
        jet pointing    &  4.6  &  2.9  &  2.6  &  2.0   \\
        L0 trigger &  4.6  &  4.5  &  4.5  &  4.4  &\\
        HLT1 trigger &  4.1  &  4.0  &  4.0  &  4.3  \\
        HLT2 trigger &  5.9  &  5.9  &  6.1  &  6.3 \\
        luminosity   &  1.7  &  1.7  &  1.7  &  1.7  \\
        \hline
        total    & 13.3  & 12.7  & 12.6  & 12.6  \\
        \hline
\end{tabular}
}
\end{table}

Several alternatives have been considered for the background
mass model, in particular with an additional exponential component,
or a component that is independent of the mass. With these models
the estimated background yield at higher mass is larger than with
the nominal background model, leading to tighter limits on the
signal. As the nominal model gives the most conservative limit, no
additional systematic uncertainty is assigned.

\section{Results}
\label{sec:Results}

The fit procedure is performed for a \vpion{} mass of 25, 35, 43 and
50\gevcc and for several values of the lifetime in between 1
and 200\ps. No significant signal is observed for any
combination of \vpion{} mass and lifetime. Upper limits are extracted
using the $\text{CL}_\text{s}$ method~\cite{Read:2002hq} with a
frequentist treatment of the nuisance parameters described above, as
implemented in the \roostats~\cite{Moneta:2010pm} package.

\begin{figure}[h]
    \centering
      \includegraphics[width=\linewidth]{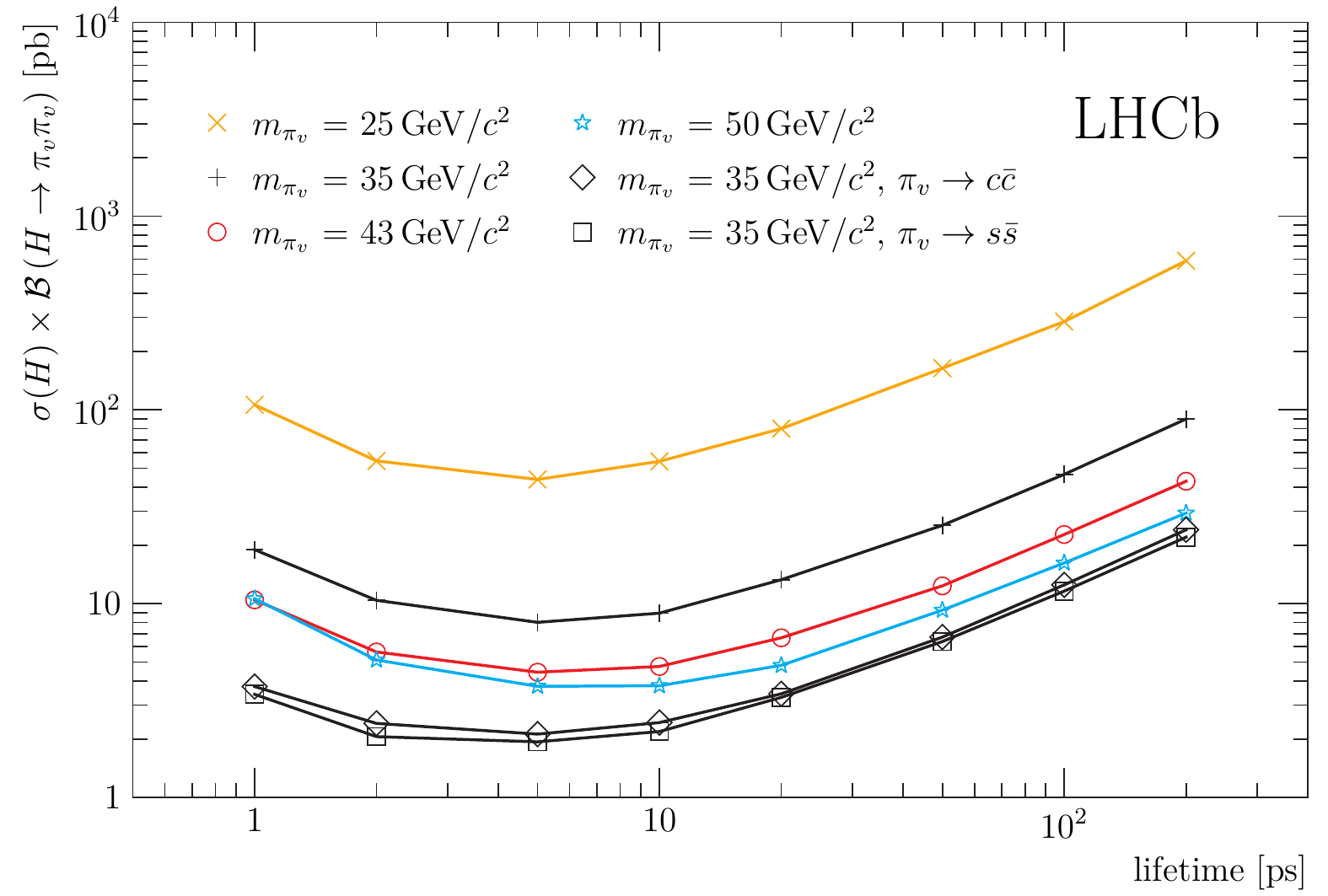}
      \caption{Observed \SI{95}{\percent} CL cross-section upper
        limits on a hidden valley
        model~\cite{Strassler:2006im,*Strassler:2006ri,*Han:2007ae}
        for various \vpion masses, as a function of \vpion{}
        lifetime. Both \vpion particles are assumed to decay
        into \bbbar, unless specified otherwise.}
  \label{fig:observedlimits}
\end{figure}

Limits are set on the Higgs production cross-section multiplied by the
branching fraction into long-lived particles $\sigma(H)\times
{\cal B}(H\to \vpion \vpion)$. In the simulation it is assumed that both \vpion{} particles decay to
the same final state. If the decay width of the
\vpion{} particle is dominated by final states other than \qqbar{}, the limits scale as $1/({\cal B}_{\qqbar}(2-{\cal
  B}_{\qqbar}))$ where ${\cal B}_{\qqbar}$ is the $\vpion\to\qqbar$
branching fraction. 
The obtained \SI{95}{\percent} CL upper limits on $\sigma(H)\times
{\cal B}(H\to \vpion \vpion)$, under the assumption of a
\SI{100}{\percent} branching fraction to \bbbar, are shown in
Table~\ref{table:observedlimits} and in
Fig.~\ref{fig:observedlimits}.
As the background decreases with
the observed di-jet invariant mass, the limits become stronger
with increasing \vpion{} mass. The sensitivity has an optimal value at
a lifetime of about 5\ps.

Additional limits are set on models with a \vpion particle decaying to
\ccbar and to \ssbar. The limits for \vpion decay to \uubar and
\ddbar are expected to be the same as for \ssbar.  The light quark
decays result in a higher displaced vertex track multiplicity, and
lighter jets, leading to a higher selection efficiency. Consequently,
the limits for decays to light quark jets are more stringent than
those for decays to $b$-quark jets.

\begin{table}[t]
  \caption{Observed \SI{95}{\percent} CL cross-section upper limits on
    $\sigma(H)\times
{\cal B}(H\to \vpion \vpion)$ (in {\pb}) on
    a hidden
    valley~\cite{Strassler:2006im,*Strassler:2006ri,*Han:2007ae} model
    for various \vpion masses and lifetimes. 
    Both \vpion particles are assumed to decay
    into \bbbar, unless specified otherwise.}
  \label{table:observedlimits}
  \centerline{
    \begin{tabular}{crrrrrrrr}
      \hline
      \vpion mass [\gevcc]    &\multicolumn{8}{c}{\vpion lifetime [\ps]} \\
      {}       & 1 & 2 & 5 & 10 & 20 & 50 &100&200\\
      \hline
      \rule{0pt}{3ex}25 &  106.3 &   54.6 &   43.8 &  54.2 &  80.0 & 164.1 &  285.7 & 588.5 \\
      35 &   19.0 &   10.4 &    8.0 &   8.9 &  13.3 &  25.4 &   46.5 &  89.8 \\
      43 &   10.5 &    5.6 &    4.4 &   4.7 &   6.7 &  12.4 &   22.7 &  42.8 \\
      50 &   10.6 &    5.1 &    3.7 &   3.8 &   4.8 &   9.3 & 16.2 & 29.3 \\
      35 (\(\vpion\to c\bar{c}\)) & 3.7 & 2.4 & 2.1 & 2.4 & 3.4 & 6.7 & 12.5 & 24.1 \\
      35 (\(\vpion\to s\bar{s}\)) & 3.4 & 2.1 & 1.9 & 2.2 & 3.3 & 6.4 & 11.6 & 22.0 \\
      \hline
    \end{tabular}
  }
\end{table}

\section{Conclusion}

A search has been presented for massive, long-lived particles in a
sample of $pp$ collisions at $\sqrt{s}=7\tev$,
corresponding to an integrated luminosity of 0.62\invfb,
collected by the LHCb experiment.
The long-lived spin-zero particles are assumed to be pair-produced in
the decay of a 120\gevcc SM Higgs, and to decay to two hadronic jets.
They appear for
instance as \vpion{} particles in hidden valley models. A single
\vpion particle is identified by a displaced vertex and two associated
jets. No significant signal for \vpion{} particles with a mass between
\num{25} and 50\gevcc and a lifetime between \num{1} and
200\ps is observed. Assuming a \SI{100}{\percent} branching fraction to
$b$-quark jets, the \SI{95}{\percent} CL upper limits on the
production cross-section $\sigma(H)\,\times\,{\cal B}(H\to \vpion \vpion)$
are in the range 4 -- 600\pb.

The results cover a region in mass and lifetime that so far
has been unexplored at the LHC. The obtained upper limits are more restrictive
than results from the Tevatron experiments in the same mass and
lifetime region. The best sensitivity is obtained for \vpion{}
particles with a lifetime of about 5\ps and a mass above
approximately 40\gevcc. The SM Higgs cross-section at
7\tev is about 17\pb~\cite{Heinemeyer:2013tqa}. The measurements in the most sensitive region exclude branching
fractions of greater than \SI{25}{\percent} for a SM Higgs boson to pair produce \vpion particles that decay to two hadronic jets.

\section*{Acknowledgements}
\noindent We express our gratitude to our colleagues in the CERN
accelerator departments for the excellent performance of the LHC. We
thank the technical and administrative staff at the LHCb
institutes. We acknowledge support from CERN and from the national
agencies: CAPES, CNPq, FAPERJ and FINEP (Brazil); NSFC (China);
CNRS/IN2P3 (France); BMBF, DFG, HGF and MPG (Germany); INFN (Italy); 
FOM and NWO (The Netherlands); MNiSW and NCN (Poland); MEN/IFA (Romania); 
MinES and FANO (Russia); MinECo (Spain); SNSF and SER (Switzerland); 
NASU (Ukraine); STFC (United Kingdom); NSF (USA).
The Tier1 computing centres are supported by IN2P3 (France), KIT and BMBF 
(Germany), INFN (Italy), NWO and SURF (The Netherlands), PIC (Spain), GridPP 
(United Kingdom).
We are indebted to the communities behind the multiple open 
source software packages on which we depend. We are also thankful for the 
computing resources and the access to software R\&D tools provided by Yandex LLC (Russia).
Individual groups or members have received support from 
EPLANET, Marie Sk\l{}odowska-Curie Actions and ERC (European Union), 
Conseil g\'{e}n\'{e}ral de Haute-Savoie, Labex ENIGMASS and OCEVU, 
R\'{e}gion Auvergne (France), RFBR (Russia), XuntaGal and GENCAT (Spain), Royal Society and Royal
Commission for the Exhibition of 1851 (United Kingdom).

\clearpage
\addcontentsline{toc}{section}{References}
\setboolean{inbibliography}{true}
\bibliographystyle{LHCb}
\bibliography{main}


\newpage

\centerline{\large\bf LHCb collaboration}
\begin{flushleft}
\small
R.~Aaij$^{41}$, 
B.~Adeva$^{37}$, 
M.~Adinolfi$^{46}$, 
A.~Affolder$^{52}$, 
Z.~Ajaltouni$^{5}$, 
S.~Akar$^{6}$, 
J.~Albrecht$^{9}$, 
F.~Alessio$^{38}$, 
M.~Alexander$^{51}$, 
S.~Ali$^{41}$
% ...

{\footnotesize \it
% ...
$ ^{5}$Clermont Universit\'{e}, Universit\'{e} Blaise Pascal, CNRS/IN2P3, LPC, Clermont-Ferrand, France\\
$ ^{6}$CPPM, Aix-Marseille Universit\'{e}, CNRS/IN2P3, Marseille, France\\
% ...
$ ^{9}$Fakult\"{a}t Physik, Technische Universit\"{a}t Dortmund, Dortmund, Germany\\
% ...
$ ^{37}$Universidad de Santiago de Compostela, Santiago de Compostela, Spain\\
$ ^{38}$European Organization for Nuclear Research (CERN), Geneva, Switzerland\\
% ...
$ ^{41}$Nikhef National Institute for Subatomic Physics, Amsterdam, The Netherlands\\
% ...
$ ^{46}$H.H. Wills Physics Laboratory, University of Bristol, Bristol, United Kingdom\\
% ...
$ ^{51}$School of Physics and Astronomy, University of Glasgow, Glasgow, United Kingdom\\
$ ^{52}$Oliver Lodge Laboratory, University of Liverpool, Liverpool, United Kingdom\\
% ...
}
\end{flushleft}

\end{document}